\documentstyle[aps,epsfig]{revtex}

\begin{document}
\draft      

%
%

\newcommand{\txt}[1]{{\mathrm #1}} 
\newcommand{\order}{\mbox{${\cal O}$}}
\newcommand{\nbody}{n-body}
\newcommand{\nplusbody}{(n+1)-body}

\newcommand{\BHO}{BHO}
\newcommand{\HERWIG}{{\tt HERWIG}}
\newcommand{\PYTHIA}{{\tt PYTHIA}}

\newcommand{\deltaS}{\delta_\txt{s}}
\newcommand{\deltaC}{\delta_\txt{c}}

\newcommand{\pT}[1]{P^T_{\mathrm #1}}
\newcommand{\pX}[1]{P^x_{\mathrm #1}}
\newcommand{\pY}[1]{P^y_{\mathrm #1}}
\newcommand{\pZ}[1]{P^z_{\mathrm #1}}
\newcommand{\vecpT}[1]{\vec{P}^T_{\mathrm #1}}
\newcommand{\mass}[1]{M_{\mathrm #1}}

\newcommand{\alphaS}{\mbox{$\alpha_\txt{S}$}}
\newcommand{\alphaEM}{ \mbox{ $\alpha_\txt{EM}$ } }
\newcommand{\NLOa}{NLO}

%
%

\title{ Incorporating next-to-leading order 
  matrix elements for hadronic diboson production 
  in showering event generators }

\author{Matt Dobbs\footnote{Electronic Mail Address: matt.dobbs@cern.ch} }
\address{
  Department of Physics and Astronomy, University of Victoria,
  P.O.\ Box 3055, Victoria, British Columbia, Canada V8W~3P6}
\date{\today}
\maketitle

\begin{abstract}

A method for incorporating information from next-to-leading order QCD
matrix elements for hadronic diboson production into showering event
generators is presented.  In the hard central region (high jet
transverse momentum) where perturbative QCD is reliable, events are
sampled according to the first order tree level matrix element. In the
soft and collinear regions next-to-leading order corrections are
approximated by calculating the differential cross section across the
phase space accessible to the parton shower using the first order
(virtual graphs included) matrix element. The parton shower then
provides an all-orders exclusive description of parton
emissions. Events generated in this way provide a physical result
across the entire jet transverse momentum spectrum, have
next-to-leading order normalization everywhere, and have positive
definite event weights.  The method is generalizable without
modification to any color singlet production process.

\end{abstract}

%
%


%
%

\section{INTRODUCTION} \label{introduction}

As high energy collider physics moves towards the TeV frontier, the
simulation of collision events at higher orders becomes increasingly
important. For the case of hadronic diboson production, higher order
corrections are largest in the regions where the physics is most
interesting. Event rate enhancements at next-to-leading order in QCD
(\NLOa) have a spoiling effect on the confidence limits for anomalous
triple gauge boson couplings, wash out the radiation zero in $W\gamma$
production~\cite{Mikaelian:1978ux} and approximate radiation zero in
$WZ$ production~\cite{Baur:1994ia}, and contribute to the background
rate for new physics channels including heavy Higgs boson searches and
Supersymmetry signatures.

Until recently, general purpose event generators have included leading
order (LO) matrix elements only.  An exclusive (meaning all of the
parton emissions are explicitly enumerated in the event record) all
orders description of multiple emissions is provided by the parton
shower approach~\cite{Sjostrand:1985xi,Marchesini:1988cf} 
%
%
which is valid in the regions of soft or collinear emissions, but is
not accurate for well-separated partons.  The shower occurs with unit
probability and so does not alter the cross section which remains
leading order.  Two of the general purpose showering Monte Carlo
programs, \HERWIG\cite{Marchesini:1992ch} and
\PYTHIA\cite{Sjostrand:1994yb}, have implemented matrix element
corrections~\cite{Seymour:1992xa,Corcella:2000gs,Miu:1999ju} to the
parton shower algorithms using the first order tree level matrix
element (the one-loop virtual graphs are not included).  The
corrections are successful in improving the agreement with data in
distribution shapes, especially at high transverse momentum.  The
normalization, however, is everywhere leading order and the overall
event rate is usually obtained by normalizing to the data or applying
a constant factor obtained from \NLOa\ calculations.


In contrast, next-to-leading (fixed) order Monte Carlo integration
programs such as the \NLOa\ calculations available for $WZ$
production~\cite{BHOWZ,Frixione:1992pj,Campbell:1999ah,Dixon:1998py}
provide a first order prediction of the cross section and a good 
%
%
description of the hardest emission. However, these
programs frequently generate events with negative weights in certain
phase space regions, do not address multiple emission, and give a poor
(indeed sometimes unphysical) description in the region of very small
jet transverse momentum where perturbative QCD is unreliable.  In this
small transverse momentum region, the effects of multiple emissions
can be resummed (first proposed by~\cite{Dokshitzer:1980hw}) giving a
better prediction of inclusive\footnote{
  An example of an inclusive quantity is the color singlet system
  transverse recoil against all parton emissions ($\pT{WZ}$ for the case
  of $WZ$ production). Inclusive means the contribution from individual
  emissions is not available, only the contribution summed over all
  emissions is known. A LO cross section is an inclusive prediction of
  the event rate. After applying a parton shower algorithm to a LO event,
  the description becomes exclusive because the enumeration of each
  parton emission is contained in the event record.}
quantities. However, the standard resummation formalism does not
describe the exclusive set of partons in the event.  In
Ref.~\cite{Han:1992sa} a resummed treatment of hadronic $Z$-pair
production $pp\rightarrow ZZ$, is given and a comparison is made with
the \PYTHIA\ showering event generator. In the region of low $Z$-pair
transverse momentum $\pT{ZZ}$, the resummed spectrum agrees reasonably
well with the results from the parton shower approach.


The combination of \NLOa\ matrix elements with parton showering
algorithms is a natural way to proceed and would provide an exclusive
Monte Carlo description of collision events. This goal has two primary
challenges: (1) A probabilistic interpretation of \NLOa\ matrix
elements is difficult to accomplish on account of the delicate
cancellations which occur between the virtual and real emission
graphs.  This generally results in event weights which may be both
positive or negative. (2) The parton shower algorithm can generate
emissions which are harder than the first (fixed order) emission
resulting in a double counting of hard emissions.  Both of these
drawbacks are accentuated in the region of small inclusive jet
transverse momentum.

Ref.~\cite{Dobbs:2001} addresses the first challenge for the special
case of hadronic $WZ$ production with leptonic decays. For each event,
a volume of low transverse momentum ($\pT{}$) parton emissions
(defined by a $\pT{}$ cutoff and assumed to be unobservable by a
detector) is integrated numerically keeping the observable quantities
(the charged lepton momenta and missing transverse momentum)
fixed. The integration volume is generated by allowing the missing
transverse momentum to arise from any combination of invisible
neutrino and low $\pT{}$ emission.  Such an integration ensures one
doesn't ``look into'' the region of small inclusive jet transverse
recoil $\pT{WZ}$, where the fixed order calculation is unreliable.
The resultant events can be interpreted probabilistically (event
weights are positive definite) and the \NLOa\ predictions for
observable distributions are preserved.  The drawback is that by
integrating the volume of jets the specific jet configuration is
lost---determining which configuration is most probable brings back
the problem of negative weights. This, and the discontinuity caused by
the $\pT{}$ cutoff, makes the subsequent application of a parton
shower algorithm difficult.

In the present study the positive features of the
Ref.~\cite{Dobbs:2001} algorithm are combined with the matrix element
corrections approach.  The phase space is divided into two
complementary regions: the region accessible to the parton shower
(``parton shower region'') and the hard central region which is not
accessible to the parton shower (``Dead Zone region'').  The treatment
of the Dead Zone region in this study follows
Ref.~\cite{Corcella:2000gs}, wherein matrix element corrections to
single vector boson production are applied within the framework of the
\HERWIG\ program.  The extension to diboson production is trivial, the
vector boson mass and momentum in Ref.~\cite{Corcella:2000gs} is
everywhere replaced by the diboson system mass and momentum.  The
parton shower region serves as a phase space volume over which
emissions can be integrated in a treatment that closely follows
Ref.~\cite{Dobbs:2001}. This provides a means of including information
from the full \NLOa\ matrix element (including one-loop graphs) such
that the normalization is \NLOa\ and the distributions reproduce the
\NLOa\ shapes in the regions where the \NLOa\ prediction is
expected to be reliable. In the low inclusive jet transverse momentum
region, where multiple emissions are important and the fixed order
prediction becomes unreliable, the distribution shapes are dominated
by the all orders parton shower provided by \HERWIG\ (though the
normalization remains \NLOa).  In this way events generated with
the algorithm described here provide an exclusive description of the
jet structure while giving a physical result across the full jet
transverse momentum spectrum and providing \NLOa\ normalization.

The method described in this paper applies to any hadronic color
singlet production process. The specific process of $pp\rightarrow
W^+Z$ production with vector boson decays to electron/muon-type
leptons at the Large Hadron Collider (LHC) energy of 14~TeV is chosen
to illustrate the method. The input parameters for this study are
$\alpha_{EM}(M_Z)=\frac{1}{128}$, $\sin^2\theta_W=0.23$,
$\alpha_s(M_Z)=0.1116$, $M_W=80.396$~GeV, $M_Z=91.187$~GeV, Cabibbo
angle $\cos\theta_\txt{Cabibbo} =0.975$ with no 3rd generation mixing,
and the factorization scale $Q$ is taken to be the hard process mass
$\sqrt{\hat{s}}$.  Vector boson branching ratios to leptons are taken
as Br($Z\rightarrow l^+ l^-$) = 3.36\%, Br($W^\pm\rightarrow
l^\pm\nu$)=10.8\%.  The {\tt CTEQ4M}~\cite{Lai:1997mg} structure
functions are used with the $b$ quark contribution taken as zero.
Matrix elements (LO and \NLOa) are evaluated with the Baur-Han-Ohnemus
(\BHO) package~\cite{BHOWZ} and \HERWIG\ is employed for the parton
cascade, hadronization, and decays.  

In the section which follows, the implementation of matrix element
corrections for diboson production in the \HERWIG\ style are presented
and a comparison is made between LO with parton shower, \NLOa, and
matrix element corrected distributions.  The method for incorporating
the full \NLOa\ matrix element in the \HERWIG\ showering event
generator is presented in Sec.~\ref{fullme} and is compared with
matrix element corrected and \NLOa\ distributions. Conclusions are
given in the final section.

%
%

\section{MATRIX ELEMENT CORRECTIONS} \label{hardme}

The parton level Born process for hadronic $WZ$ production is
$q\bar{q}'\rightarrow WZ$. The kinematics describing this process are
given the label `\nbody' and the label `\nplusbody' is reserved for
the first order tree processes $q\bar{q}'\rightarrow WZg$,
$qg\rightarrow WZq'$, and $g\bar{q}\rightarrow WZ\bar{q}'$.

Given the kinematics of an \nbody\ event as input, a showering Monte
Carlo program uses backwards evolution to trace the hard scattering
partons back from the hard vertex into the original incoming hadrons
while keeping track of the distribution of emitted partons (refer to
Ref.~\cite{QCD96} for a good review).  For the case of the \HERWIG\ 
program, the phase space accessible to the parton shower (the parton
shower region) is a subset of the full \nplusbody\ phase space. The
boundary of the Dead Zone region inaccessible to the \HERWIG\ parton
shower is defined by~\cite{Corcella:2000gs}
\begin{equation} \label{shat_limit}
  \frac{7-\sqrt{17}}{2}M^2 < \hat{s} < s
\end{equation}
and
\begin{equation} \label{cos_limit}
  |\cos \hat{\theta}_j | < 1 
  - \frac{ 3 - \sqrt{ 1 + 8 \frac{M^2}{\hat{s}} } }
  { \frac{\hat{s}}{M^2} -1 }
\end{equation}
where $M$ is the mass of the color singlet object ($WZ$ system in this
case), $\sqrt{\hat{s}}$ is the mass of the color singlet and parton
emission system ($WZj$), $\sqrt{s}$ is the machine energy, and
$\hat{\theta}_j$ is the angle of the emission with respect to the beam
in the center of mass frame. The partition of phase space in the
$\hat{s}/M^2$ vs. $\cos\hat{\theta}_j$ plane is shown in
Figure~\ref{shat_costheta_plane}. The parton shower and Dead Zone
regions are precisely complementary, so there can be no double
counting when sampling the regions separately. The Dead Zone is the
region of hard central emissions and can be populated by sampling the
first order tree level matrix element within the boundaries defined in
Eqs.~\ref{shat_limit}~and~\ref{cos_limit}. Adding the Dead Zone phase
space in this manner to the showering event generator simulation is
termed `hard matrix element corrections'. For single boson production
the improvement in the boson recoil transverse momentum
spectrum~\cite{Corcella:2000gs} is remarkable, particularly at high
$\pT{}$.

\begin{figure}
  \mbox{\epsfig{file=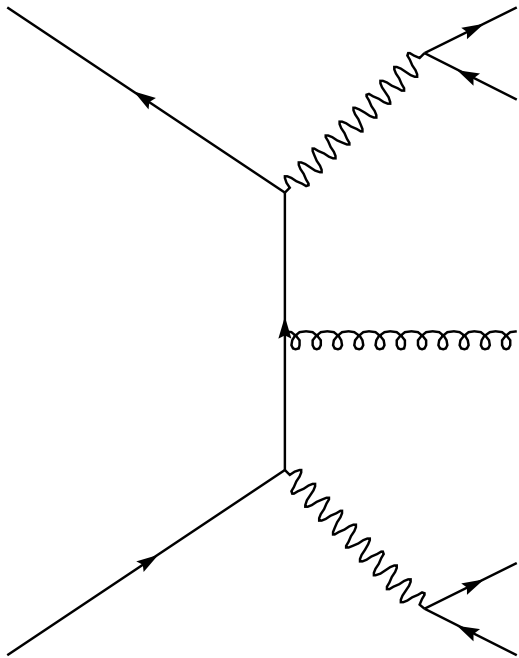,width=8.6cm}}
  \caption{ 
    Example of a first order $WZ$ production graph with a gluon
    anchored to an internal line.
    }
  \label{internal_line}
\end{figure}

`Soft matrix element
corrections'~\cite{Seymour:1992xa,Corcella:2000gs} refers to
correcting emissions generated by the parton shower algorithm using
the first order tree level matrix element.  Since the \HERWIG\ parton
shower ordering variable is an energy-weighted angle which does not
necessarily imply ordering in transverse momentum, it is necessary to
apply the correction to every emission which has the highest $\pT{}$
so far. The soft corrections are described for the single boson
production case in Ref.\cite{Corcella:2000gs} and were noted to have a
small effect in comparison to the hard matrix element corrections for
single boson production. The main advantage of the soft corrections is
in providing for a smooth transition between the parton shower and
Dead Zone regions---though in practice it is difficult to find
observables sensitive to this transition.
The primary difference with regards to soft corrections for diboson
production as compared to the single boson case are the real emission
graphs involving a gluon anchored to an internal line, such as the one
shown in Figure~\ref{internal_line}.  These graphs are absent for the
single boson case and are in no way approximated by the parton shower
which traces only external lines. In contrast to the other emission
graphs, these graph do not have any collinear singularities.  This
means that their relative contribution to the cross section is largest
in the central regions (since other emission graphs dominate in the
collinear regions). Thus, the relative effect of these graphs in the
parton shower region is small, and it is more important to account for
them in the Dead Zone region where the physics of these graphs may
become important.

The present study uses the soft matrix element corrections already
implemented in \HERWIG\ for single boson production. Since the
emission cross section enters only as a ratio to the Born cross
section, the \HERWIG\ implementation accounts for the first order
emission from an external line of Feynman graphs for color singlet
production to good approximation, but does not account for emissions
from an internal line. Implementing the soft corrections for the
specific case of $WZ$ production would require intrusive changes to
the \HERWIG\ parton shower code. By accepting this approximation, the
\HERWIG\ code is kept unmodified and modular.  Nevertheless it is
recognized that explicit diboson soft matrix element corrections would
be a useful extension to this study.

The \HERWIG\ soft matrix element corrections (which are approximate
for the diboson case) have very little impact on the distributions
presented in this paper. It is expected that the exact soft matrix
element corrections (i.e.\ explicitly including all diboson graphs)
would not change the situation in an appreciable way.

\subsection{ Matrix element corrected results }

\begin{figure}
  \mbox{\epsfig{file=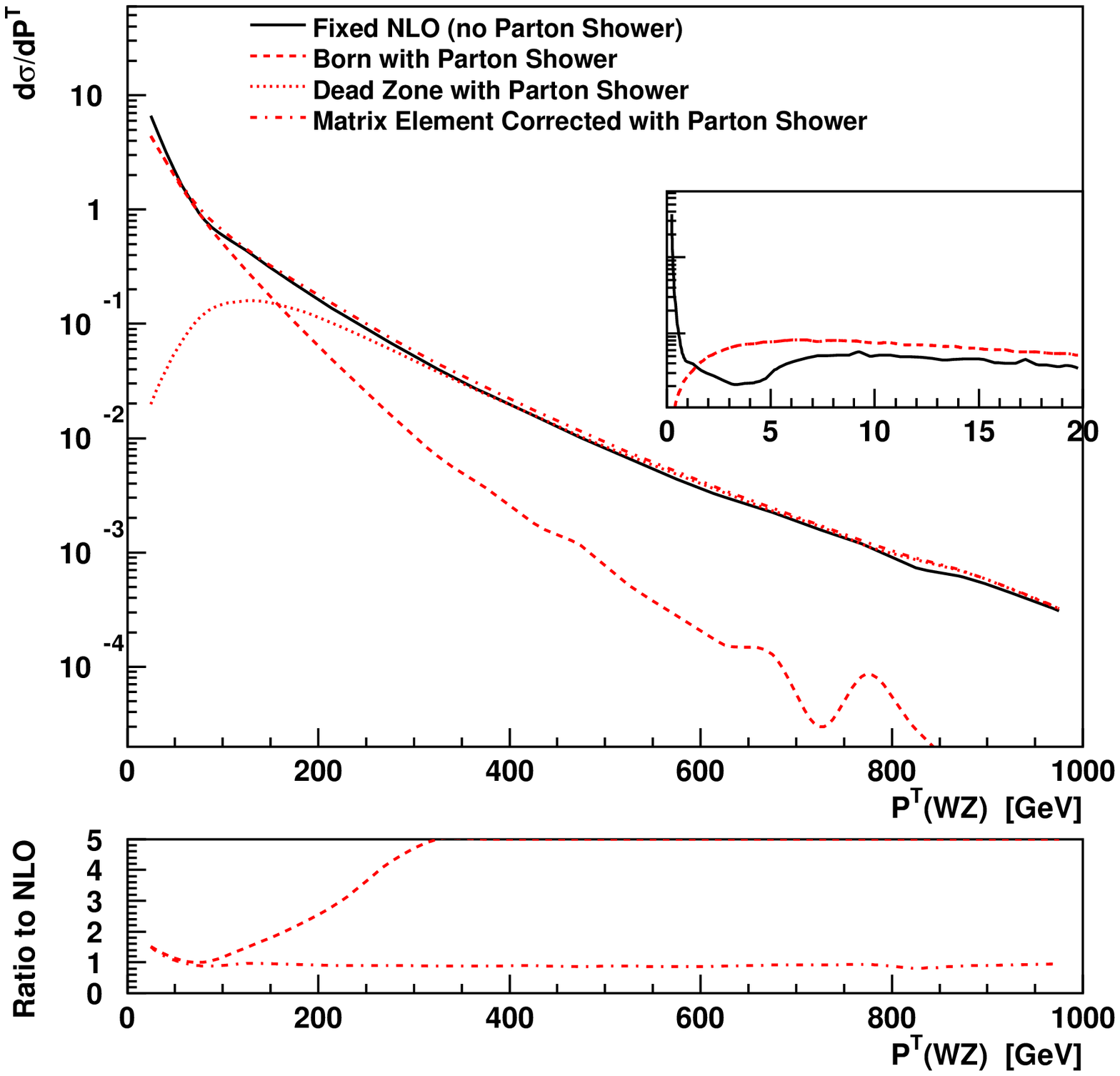,width=8.6cm}}
  \caption{ 
    The transverse momentum distribution of the $W^+Z$ system (inclusive
    recoil against emissions) $\pT{WZ}$ is compared at \NLOa, LO with
    parton shower, first order tree level in the Dead Zone, and at LO
    with soft and hard matrix element corrections.  An enhanced view
    of the first low $\pT{WZ}$ bin is shown in the inset and the ratio
    of the \NLOa\ distribution to the LO and matrix element corrected
    distributions is shown at bottom.  }
  \label{PTWZhardme}
\end{figure}

LO and matrix element corrected distributions for hadronic $WZ$
production are obtained by
implementing the Born level \nbody\ process and tree level \nplusbody\ 
process (restricted to the Dead Zone) as external processes in
\HERWIG\ 
%
%
using the \BHO\ matrix elements.  For diboson production, the
distribution corresponding to the single vector boson production
transverse recoil studied in Ref.~\cite{Corcella:2000gs} is the diboson
system recoil $\pT{WZ}$.  A comparison to the \NLOa\ result for this
distribution is shown in Figure~\ref{PTWZhardme} for the LO with
parton shower, first order tree level in the Dead Zone, and matrix
element corrected results.
Away from the low $\pT{}$ region, the shape of the
matrix element corrected result agrees well with the \NLOa\ prediction
from the stand-alone BHO integration package, demonstrating the
effectiveness of hard matrix element corrections.  There is no
contribution from the Dead Zone in the very low $\pT{WZ}$ region, so
the Born and matrix element corrected results coincide in this region
(Figure~\ref{PTWZhardme}, inset).  The soft matrix element corrected
LO result is indistinguishable from the LO curve and is not shown.  

\begin{figure}
  \mbox{\epsfig{file=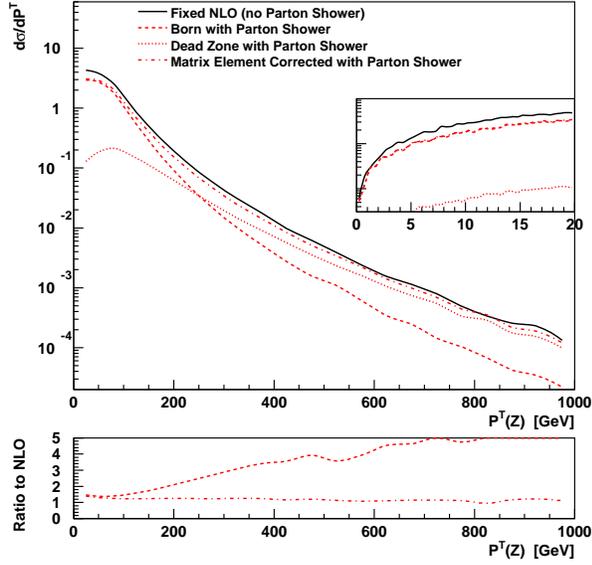,width=8.6cm}}
  \caption{
    The transverse momentum distribution of the $Z$-boson $\pT{Z}$ is
    compared at \NLOa, LO with parton shower, first order tree level
    in the Dead Zone, and at LO with soft and hard matrix element
    corrections.  An enhanced view of the first low $\pT{Z}$ bin is
    shown in the inset and the ratio of the \NLOa\ distribution to the
    LO and matrix element corrected distributions is shown at bottom.
    }
  \label{PTZhardme}
\end{figure}

It is more common in the literature to study the $\pT{Z}$
distribution, for example in probing the triple gauge boson vertex for
anomalous couplings. This distribution is also sensitive to hard
emissions in the Dead Zone, as shown in Figure~\ref{PTZhardme} where,
as for the $\pT{WV}$ spectrum, the shape of the matrix element
corrected $\pT{Z}$ distribution is a good approximation to the \NLOa\ 
result.

Drawbacks in the matrix element corrections approach and
fixed order calculation are evident in Figure~\ref{PTWZhardme}. The
\NLOa\ calculation is not reliable in the very low $\pT{}$ region
shown in the inset of Figure~\ref{PTWZhardme}. The \NLOa\ prediction
shows a dip just below 5~GeV and a very large contribution at zero.
These features are not physical and reflect the specific
choices of the regularization scheme used in the matrix elements. A
different choice of scheme or scheme parameters would yield a very
different result in this region.\footnote{ 
  The \BHO\ matrix elements employ the two parameter phase space slicing
  method~\cite{Baer:1989jg} as regularization scheme. The specific
  choice of cutoff parameters used for Figure~\ref{PTWZhardme} are
  large ($\deltaS=0.05,~\deltaC=0.01$ see Ref.\cite{BHOWZ} for a
  definition of the cutoffs). A smaller choice of parameters would
  move events out of the zero histogram bin of the
  Figure~\ref{PTWZhardme} inset towards regions of higher $\pT{}$,
  until eventually the zero bin would become negative.  The use of the
  subtraction method~\cite{Ellis:1981wv} for the $\pT{WZ}$
  distribution in the inset of Figure~\ref{PTWZhardme} also
  yields an unphysical negative first histogram bin. This effect serves
  as a reminder that fixed order perturbation theory is not well
  suited for predictions of small inclusive $\pT{}$.  }

\begin{figure}
  \mbox{\epsfig{file=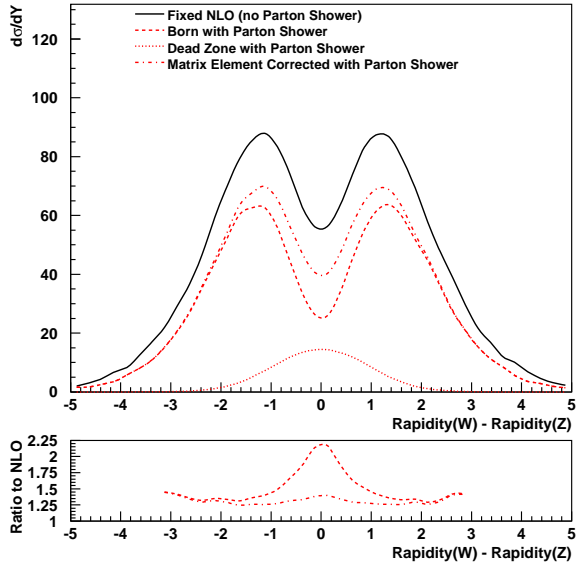,width=8.6cm}}
  \caption{
    The rapidity separation distribution of the $W^+$ and $Z$-bosons is
    compared at \NLOa, LO with parton shower, first order tree level
    in the Dead Zone, and at LO with soft and hard matrix element
    corrections.  The dip in the zero separation region is the
    approximate radiation zero. The ratio of the \NLOa\ distribution
    to the LO and matrix element corrected distributions is shown at
    bottom.  }
  \label{YWZhardme}
\end{figure}

A disadvantage of the matrix element corrected distributions is the LO
normalization in the parton shower region.  Since the total matrix
element corrected cross section is the sum of the Born level and the
integrated Dead Zone cross sections, it depends explicitly on the Dead
Zone boundary defined by Eqs.~\ref{shat_limit}~and~\ref{cos_limit}. A
change in the boundary would result in a change in the total cross section.
The use of \NLOa\ calculations everywhere would remove this dependence and
provide a more accurate prediction of the total cross section,
particularly for multi-TeV colliders where the contribution from the
gluon (anti-)quark initial state is substantial.  A second drawback of
the matrix element corrections is the inability to model \NLOa\ shape
effects in the parton shower region.  The approximate
radiation zero, which is partially filled in by \NLOa\ corrections, is
an example of a distribution which is sensitive to these effects.
Observation of this as yet unseen phenomenon will be a goal of future
hadron colliders. This distribution is shown in
Figure~\ref{YWZhardme}, together with the ratio (or $k$-factors) of
the matrix element corrected and Born distributions with respect to
the \NLOa\ distributions. The $k$-factors are largest in the central
radiation zero region and the matrix element corrections are
successful in greatly reducing this effect. However, a small variation
in $k$-factors in the zero rapidity separation region remains after
applying the matrix element corrections.

%
%

\section{INCORPORATING THE FULL \NLOa\ MATRIX ELEMENT} \label{fullme}

In the Dead Zone the square of the real emission graphs
(${\cal{M}}_{\mathrm real~emission}^2$) is the only contribution to
the cross section and the divergent regions of the real emission phase
space are excluded.  Outside of the Dead Zone the parton shower is
better suited at predicting event shapes and exclusive distributions,
though it provides no information on the normalization which in
showering event generators is normally derived from the LO cross
section only. In this region it is preferable to obtain the
normalization from the full \NLOa\ cross section while preserving the
exclusive event structure from the parton shower.

\begin{figure}
  \mbox{\epsfig{file=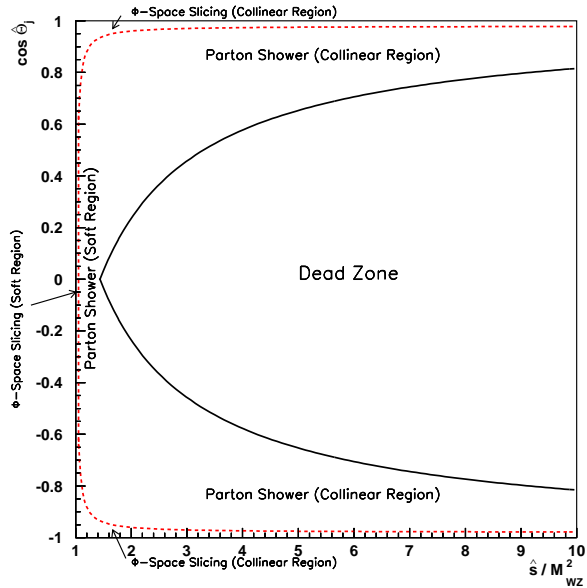,width=8.6cm}}
  \caption{ 
    The $\hat{s}/M^2$~vs.~$\cos\hat{\theta}_j$ plane is shown with the
    boundary between the complimentary parton shower and Dead Zone
    regions denoted with a solid line. The phase space slicing region
    is a subset of the parton shower region and is denoted with a
    dashed line (for the specific phase space slicing parameters
    $\deltaS=0.05,~\deltaC=0.01$).
 }
  \label{shat_costheta_plane}
\end{figure}

In the parton shower region the square of the Born graphs,
the interference of the Born graphs with the one-loop graphs, and the
square of the real emissions graphs contribute at \NLOa,
\begin{equation} \label{NLO_ME}
        {\cal{M}}_{\mathrm NLO}^2 ~=~ 
        {\cal{M}}_{\mathrm Born}^2 ~+~
        {\cal{M}}_{\mathrm Born} ~\otimes~ {\cal{M}}_{\mathrm one~loop} ~+~
        {\cal{M}}_{\mathrm real~emission}^2.
\end{equation}
The first two terms on the right hand side of Eq.~\ref{NLO_ME} are
described by \nbody\ kinematics and occupy a point at
$(\hat{s}/M^2,\cos\hat{\theta}_j)=(1,0)$ of
Figure~\ref{shat_costheta_plane}. The last term is described by
\nplusbody\ kinematics and may populate the entire $\hat{s}/M^2$ vs.
$\cos\hat{\theta}_j$ plane.  A regularization scheme is necessary to
handle the soft, collinear, and ultraviolet divergences which appear
when any of the first order contributions are treated alone. The \BHO\ 
matrix elements use the phase space slicing method~\cite{Baer:1989jg}
(PSS) wherein a small portion (denoted on
Figure~\ref{shat_costheta_plane}) of the \nplusbody\ phase space
(defined by soft and collinear cutoff parameters $\deltaS,~\deltaC$)
is included in the \nbody\ contribution by means of the soft gluon and
leading pole approximations.

One way of achieving \NLOa\ normalization while gaining sensitivity to
the effect of \NLOa\ corrections on the $WZ$ system configuration is
to assign an event weight for the hard process which is the average
\NLOa\ cross section (including virtual graphs) across the entire
region accessible to the parton shower for a particular choice of the
$WZ$ system configuration.  Since all of the divergent regions are
contained within the parton shower region, this number is well behaved
and positive definite---but is difficult to calculate analytically.

This average cross section can be obtained by means of a 2nd stage
integration as outlined in Ref.~\cite{Dobbs:2001}.  It is also noted
that it is not necessary to obtain this number with a high degree of
accuracy---so long as the average of many such integrations (i.e.\ 
events) converges to the correct cross section, which is automatic
using the Monte Carlo method.

For the present algorithm the jet volume has a simple definition
(Eqs.~\ref{shat_limit}~and~\ref{cos_limit}) with
the convenient feature that the phase space slicing method jet volume
is a subset of this volume. This means the phase space
slicing method can be used to achieve a large portion of the jet
volume integration and that the result is not sensitive to the
specific choice of PSS cutoff parameters which specify the PSS
boundary. As such, the maximal PSS cutoffs (which remain within the range
of validity of the soft gluon and leading pole approximations) may be
used ($\deltaS=0.05,~\deltaC=0.01$ have been used, see
Ref.\cite{BHOWZ} for a definition of the cutoffs).  This is a
significant advantage because the subset occupied by the PSS volume is
the region where delicate cancellations between the \nbody\ and
\nplusbody\ graphs occur and also where the cross section is peaked.

The 2nd stage integration, which is performed once for each event in
the parton shower region, is accomplished as follows. A $WZ$ event
with \nbody\ kinematics is generated (8 degrees of freedom specify the
\nbody\ configuration assuming the vector bosons are taken on shell:
e.g.\ the $WZ$ system mass $\mass{WZ}$, $WZ$ system rapidity
$y_{\mathrm WZ}$, the 2 angles defining the vector boson production,
and 2 angles for each of the vector boson decays) and defines the $WZ$
system configuration and boost. For a LO showering Monte Carlo event,
the Born level cross section would be evaluated and constitute the
event weight. Instead, the aim is to evaluate the average cross
section of the phase space into which the parton shower may evolve the event.

The \NLOa\ \nbody\ cross section is calculated using the \BHO\ matrix
element which requires a further 4 degrees of freedom to specify the
soft and collinear corrections arising in the PSS method. Thus by
integrating over these 4 degrees of freedom the average \nbody\ weight
$\left< d\sigma^{\mathrm \nbody} \right>$ is obtained and accounts for
the region inside the PSS boundary (on the left side of the dashed
line) in Figure~\ref{shat_costheta_plane}.  This number may be
negative.  For the integration of the remainder of the parton shower
region an emission is explicitly sampled using 3 additional degrees of
freedom (e.g.\ the $WZj$ system mass and the 2 angles specifying the
emission direction).  These 11 degrees of freedom (3 plus 8 from the
\nbody\ configuration) specify the \nplusbody\ kinematics.  It is
necessary to use the appropriate overall boost, so as to span the
identical phase space the \HERWIG\ parton shower is able to reach from
the \nbody\ kinematics.  \HERWIG\ applies a conformal boost to the
hard process which keeps the hard process mass and rapidity fixed. For
the \nplusbody\ case the $WZj$ system rapidity is thus
\begin{equation}
  y_{\mathrm \nplusbody} ~=~ \ln \left( \frac
    {\sqrt{ {\mass{WV}}^2 + {\pT{WZ}}^{\star~2} }}
    { {E_{\mathrm WZ}}^\star + {\pZ{WZ}}^\star } \right)
    ~+~ y_{\mathrm \nbody}
\end{equation}
%
where $y_{\mathrm \nbody}$ is identified as the $WZ$ system rapidity
and ${P_{\mathrm WZ}}^\star$ is the $WZ$ system 4-vector in the $WZj$
frame. The Jacobian for the transformation $y_{\mathrm \nbody}
\rightarrow y_{\mathrm \nplusbody}$ is unity.  
%
By sampling the 3 degrees of freedom which specify the expansion from
\nbody\ to \nplusbody\ kinematics and evaluating the \NLOa\ matrix
element each time (always a positive number), an integration over the
remainder of the parton shower region is achieved giving the average
\nplusbody\ weight $\left< d\sigma^{\mathrm \nplusbody} \right>$.  

The cross section for the specific choice of $WZ$ configuration
averaged over the parton shower region at \NLOa\ is
\begin{equation}\label{EventWeight}
  d\sigma^{\mathrm parton~shower~region}~=~
                \left<d\sigma^{\mathrm \nbody}\right> 
            ~+~ \left<d\sigma^{\mathrm \nplusbody}\right>
\end{equation}
and constitutes the event weight.  The input to the parton shower
algorithm is the \nbody\ kinematic configuration. The parton shower
algorithm (with guidance from the soft matrix element corrections)
determines which exclusive emission structure is chosen for the event.

Event generation for the \NLOa\ $WZ$ process consists of generating a
sample of events in the parton shower region using the 2nd stage
integration method discussed above and a second sample of events in
the Dead Zone, as already discussed in Sec.~\ref{hardme}.  The event
weights for both types of events are positive numbers (unlike for the
\NLOa\ integration packages which produces events with both positive and
negative weights) so the usual event generator strategy of
hit-and-miss may be used to obtain unweighted event samples.

\subsection{Results}

The cross sections for the regions of interest are enumerated in
Table~\ref{cross_sections} for $W^+Z$ production at LHC with decays to
electron/muon-type leptons. The total Born level cross section is
almost a ($k$-)factor 1.5 down from the \NLOa\ result. The Dead Zone
accounts for about 8\% of the total cross section. The sum of the
integrated parton shower region cross section and the Dead Zone cross
section agrees with the total \NLOa\ cross section which serves as a
cross check that the parton shower region contribution is correctly
evaluated with the 2nd stage integration.

\begin{figure}
  \mbox{\epsfig{file=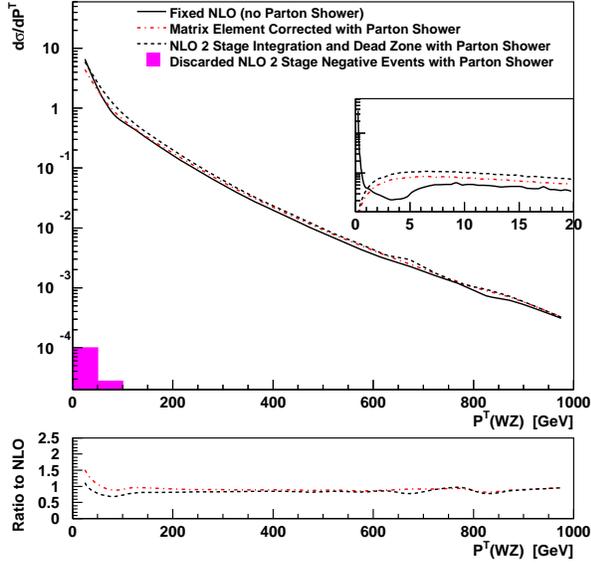,width=8.6cm}}
  \caption{
    The transverse momentum distribution of the $W^+Z$ system (inclusive
    recoil against emissions) $\pT{WZ}$ is compared at \NLOa, at LO
    with soft and hard matrix element corrections, and using the 2nd
    stage integration (which also employs soft and hard matrix element
    corrections).  An enhanced view of the first low $\pT{WZ}$ bin is
    shown in the inset and the ratio of the \NLOa\ distribution to the
    matrix element corrected and 2nd stage integration distributions
    is shown at bottom.  }
  \label{PTWZfullme}
\end{figure}

In Figure~\ref{PTWZfullme} the transverse recoil of the $WZ$ system
is shown for this 2nd stage integration method (which also includes
matrix element corrections).  The normalization is the same as the
\NLOa\ normalization by construction. The \NLOa\ and matrix element
corrected results from Figure~\ref{PTWZhardme} are superimposed.  In
the region of moderate to high $\pT{WZ}$ the result agrees well with
the \NLOa\ prediction. In the region of smaller $\pT{WZ}$, events with
low transverse momentum receive extra recoil from multiple emissions
and are pushed towards slightly higher $\pT{WZ}$. At very small
$\pT{WZ}$ (inset) the shape agrees very well with the parton shower
distribution and goes smoothly to zero with no dependence on the
\NLOa\ regularization scheme. The ratio to the fixed \NLOa\ 
distribution is slightly flatter (Fig.~\ref{PTWZfullme}, bottom) than
for the matrix element corrected distribution.

A second distribution which is sensitive to the region of small
inclusive jet $\pT{}$ is the azimuthal separation of the vector bosons
shown in Figure~\ref{PHIfullme}. As the separation approaches $\pi$,
the inclusive jet $\pT{}$ goes to zero and the fixed order prediction
becomes unreliable, as evidenced by the sharp increase in differential
cross section very near $\phi(W)-\phi(Z)=\pi$. The 2nd stage
integration and matrix element corrected results do not suffer from
this discontinuity and give a smooth, physical prediction across the
full range which is higher than the \NLOa\ result away from
$\phi(W)-\phi(Z)=\pi$. This is an example of a distribution where
the \NLOa\ prediction is not accurate.

The $Z$ transverse momentum is shown in Figure~\ref{PTZfullme} for the
2nd stage integration method, the matrix element corrections, and the
\NLOa\ calculation. The 2nd stage integration is in good agreement
with the \NLOa\ prediction throughout, while the matrix element
corrections suffer from normalization problems. The \NLOa\ calculation
is accurate in the low $\pT{Z}$ region for this observable since an
integration over a large range of inclusive jet $\pT{}$ is automatic
for the low $\pT{Z}$ bins.

The rapidity separation of the vector bosons is shown in
Figures~\ref{YWZfullme} for the 2nd stage integration method, the
matrix element corrections, and the \NLOa\ calculation. The 2nd stage
integration is in good agreement with the \NLOa\ prediction
throughout, with a very slight enhancement visible in the radiation
zero region due to multiple emissions.  The ratio of the 2nd stage
integration distribution to the \NLOa\ result is unity and almost constant
throughout, whereas the matrix element corrected ratio is about 1.3
with a small variation in the zero separation region. The 2nd stage
integration has accounted for the effects of the first order
corrections on the $WZ$ system configuration (including the
separation in rapidity) and so this small variation is not present in that
distribution.

\subsection{ Complications due to negative weights }

\begin{figure}
  \mbox{\epsfig{file=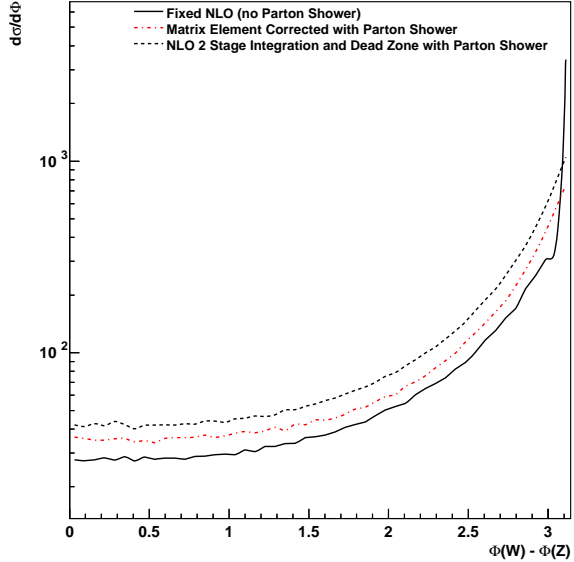,width=8.6cm}}
  \caption{
    The azimuthal vector boson separation $\phi(W^+)-\phi(Z)$
    distribution is compared at \NLOa, at LO with soft and hard
    matrix element corrections, and using the 2nd stage integration
    (which also employs soft and hard matrix element corrections).}
  \label{PHIfullme}
\end{figure}

\begin{figure}
  \mbox{\epsfig{file=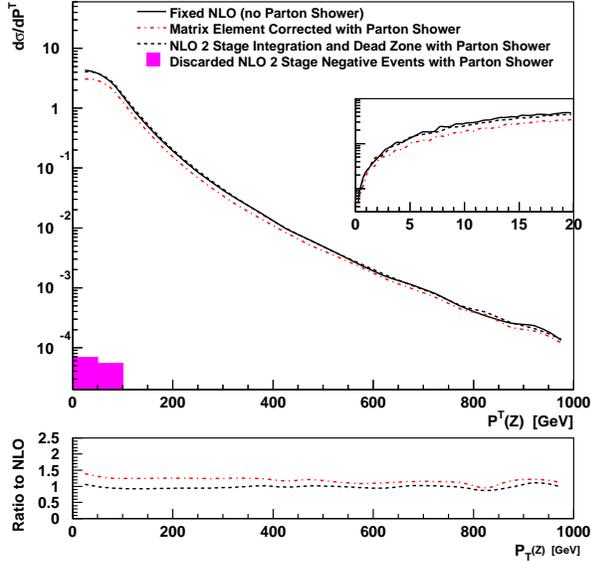,width=8.6cm}}
  \caption{
    The transverse momentum distribution of the $Z$-boson $\pT{Z}$ is
    compared at \NLOa, at LO with soft and hard matrix element
    corrections, and using the 2nd stage integration (which also
    employs soft and hard matrix element corrections).  An enhanced
    view of the first low $\pT{Z}$ bin is shown in the inset and the
    ratio of the \NLOa\ distribution to the matrix element corrected
    and 2nd stage integration distributions is shown at bottom.  }
  \label{PTZfullme}
\end{figure}

\begin{figure}
  \mbox{\epsfig{file=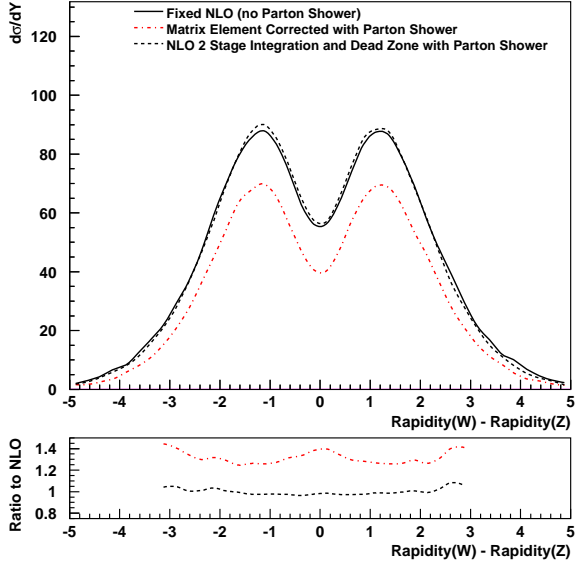,width=8.6cm}}
  \caption{
    The rapidity separation distribution of the $W^+$ and $Z$-bosons is
    compared at \NLOa, at LO with soft and hard matrix element
    corrections, and using the 2nd stage integration (which also
    employs soft and hard matrix element corrections).  The dip in the
    zero separation region is the approximate radiation zero. The
    ratio of the \NLOa\ distribution to the matrix element corrected
    and 2nd stage integration distributions is shown at bottom. }
  \label{YWZfullme}
\end{figure}

For all of the 2nd stage integration method distributions
(Figures~\ref{PTWZfullme}-\ref{YWZfullme}) the integration over the
parton shower region (performed once per event) is accomplished with
50~samples of the 3~degrees of freedom which define the expansion to
\nplusbody\ phase space and 10~samples of the 4~degrees of freedom
which define the \nbody\ PSS corrections.  The 2nd stage integration
increases the computer time required to generate one weighted event
with the \HERWIG\ showering event generator by a factor 1.3.\footnote{
  The computation time will increase significantly when generating
  unweighted events. }
This is due mostly to the integration, but also to the added number of
mathematical operations necessary to evaluate the \NLOa\ matrix
element as compared to the LO one.

Since the \nbody\ contribution can be negative, the total event weight
(Eq.~\ref{EventWeight}) can be within statistical precision of zero if
the integration statistics (number of samples) are small enough and the
cancellations (between \nbody\ and \nplusbody\ contributions) are
large enough. This means that negative events can very occasionally
occur (a discussion of 2nd stage integration negative weighted events is
given in Ref.~\cite{Dobbs:2001}\footnote{
  The lepton correlation effect discussed in Ref.~\cite{Dobbs:2001}
  also occurs in the present study, but the effect is small.}
). Negative events of this type are simply discarded.  The bias
introduced by discarding these negative events is easy to evaluate.
The absolute value of the discarded negative weight events is
superimposed on Figures~\ref{PTWZfullme}-\ref{YWZfullme} as a shaded
histogram. These histograms are 5 orders of magnitude below the 2nd
stage integration distributions (too small to resolve on
Figures~\ref{PHIfullme}~and~\ref{YWZfullme}) and are largest in the
region where the differential cross section is also largest, and so
the effect of these negative events is everywhere negligible.

It is possible to use as few as one sample of each type (\nbody\ and
\nplusbody) for the 2nd stage `integration'. The effect of
decreasing the samples in this way is to increase the frequency of
negative weight events---in this case the bias increases to 1\%, which
represents a maximum ceiling for the bias.

Finally it is noted that an adaptation of the PSS method may eliminate
the negative event weights entirely allowing for the unbiased use of a
single sample to evaluate the parton shower region cross section.  In
Ref.~\cite{Potter:2000an} a hybrid~\cite{Glover:1995vz} of the
subtraction and the one parameter PSS regularization schemes is
investigated. For each \nbody\ configuration the PSS cutoff parameter
which gives a zero \nbody\ contribution is calculated---in this manner
the virtual and real emission cancellations are enforced and all of
the phase space is partitioned to be \nplusbody. One concern with this
method is that the minimum jet energy is coupled to the \nbody\ 
momentum configuration. However, there can be no bias for any
observable which is insensitive to the specific PSS boundary for a
given \nbody\ configuration. So long as the hybrid PSS region is
everywhere contained within the parton shower region (which is
expected to be the case), the average cross section of the parton
shower region is one such number---and so this method could be
employed to evaluate the 2nd stage integration. In this manner the
\NLOa\ cross section across the parton shower region
(Eq.~\ref{EventWeight}) can be represented with a single sample, which
with this hybrid method is always a positive number. The weights of
many such events will converge to the correct cross section, and event
generation would proceed with a single sampling of the phase space, as
it normally does for LO showering event generators.  The important
difference is that in the parton shower region information about the
parton emission would be discarded and the \nbody\ kinematic
configuration would be used as input for the parton shower, which
eliminates the possibility of the jet energy coupling since, as
described previously, the parton shower (guided by the soft matrix
element corrections) would determine which exclusive event structure
is chosen for the event.

%
%

\section{CONCLUSIONS} \label{conclusions}

Matrix element corrections for hadronic diboson production have been
implemented following the procedure outlined in
Ref.~\cite{Corcella:2000gs} and are
interfaced to the \HERWIG\ showering event generator. The corrections
are significant in the high-$\pT{}$ regions where the agreement in
shape with \NLOa\ calculations is significantly improved.

A new method for incorporating the full \NLOa\ matrix element is
presented. The inclusive recoil against emissions is dominated by the
parton shower for events in the soft and collinear regions where the
parton shower is expected to give a good description of the physics,
and is dominated by the first order matrix element for events in
the hard central region where perturbation theory provides a reliable
prediction. The normalization is everywhere \NLOa.  Events generated
with this method have positive definite event weight, such that
unweighted events can be easily produced by means of Monte Carlo
hit-and-miss. The method is generalizable to any color singlet
production process.

The numerical implementation of the method is straight-forward
providing a \NLOa\ matrix element (employing the PSS regularization scheme)
is available. 



The difference in distribution shapes between the matrix element
correction approach and the 2nd stage integration incorporating the
full \NLOa\ matrix element is small---though noticeable for example in
distributions such as the diboson rapidity separation. The 2nd stage
integration approach has the further advantage of providing \NLOa\
normalization.  

Diboson event samples from hadron colliders to date are small and
insensitive to QCD corrections, so a meaningful comparison to data
will have to await Tevatron Run~II and LHC.

%
%

\section*{ACKNOWLEDGMENTS} \label{acknowledgments}

The author would like to thank the ATLAS Collaboration and in
particular I.~Hinchliffe, M.~Lefebvre, and J.~B.~Hansen.
I am grateful to U.~Baur for providing
the matrix elements, encouragement, and informative correspondence.  
I benefited from discussions with G.~Corcella, M.~Mangano,
G.~Ridolfi, M.~Seymour, and B.~Webber.
I thank L.~Bourhis, A.~Signer, and J.~Stirling for organizing
and hosting the 2000 IPPP Workshop on Matrix Elements and Parton
Showers in Durham.
%
%
This work has been supported by the Natural Sciences and Engineering
Research Council of Canada.

%
%

%
%

%
%

%
%

%
%

%
%

%
%

%
%

%
%

%
%

%
%

\begin{table}
\caption{
  The integrated cross section for $W^+Z$ production with decays to
  electron/muon-type leptons at LHC evaluated using the \BHO\ matrix
  elements is tabulated. The ratio (or $k$-factor) of the \NLOa\
  calculation to the Born is about 1.48, which improves to 1.32 when the Dead
  Zone region is added to the Born result. The ratio for the \NLOa\
  calculation to the sum of the 2nd stage integration and the Dead Zone 
  is unity by construction.
}
\label{cross_sections}
\begin{tabular}{ l l }
  Born                            & 288.2$\pm$0.3 fb \\
  Dead Zone region (\NLOa)        & 34.43$\pm$0.02 fb \\
  2nd stage integration over parton shower region (\NLOa)
                                  & 391.8$\pm$0.3 fb \\
  \NLOa\ (all phase space included) & 426.0$\pm$0.5 fb \\
\end{tabular}
\end{table}

%
%

\end{document}